\documentclass{mem}
\usepackage{natbib}\usepackage{txfonts}\usepackage{balance}
\usepackage{graphicx}
\idline{75}{282}
\begin{document}
\def\teff{$T\rm_{eff }$}
\def\kms{$\mathrm {km s}^{-1}$}

\title{Lithium and $p$-capture elements in globular clusters: implications for multiple population scenarios
}

   \subtitle{}

\author{
V. D'Orazi\inst{1,2} 
\and
R. Gratton\inst{1}
          }

\institute{
INAF -- Osservatorio Astronomico di Padova, vicolo dell'Osservatorio 5, 
35122, Padova, Italy.
\email{valentina.dorazi@inaf.it}
\and
School of Physics and Astronomy, Monash University, Clayton, VIC 3800, Melbourne, Australia.
}

\authorrunning{D'Orazi \& Gratton }

\titlerunning{Lithium and multiple populations in GCs}

\abstract{In the multiple population framework, a number of studies have been 
accomplished in order to explore the behaviour of lithium with proton-capture element abundances (e.g., Na, O, Al) in globular cluster stars.
Lithium offers perhaps one of the most severe constraints on the stellar source of internal pollution in these complex systems. Given its vulnerability,
we expect that material processed via the hot H-burning, re-cycled in the formation of the subsequent generation(s), is free of Li. However, Nature breaks our expectations.  
 In this contribution we will review the current status of this field, by examining the controversial, surprising results and implications.
}
\maketitle{}

\section{Introduction}
A few decades of extensive and dedicated observations have unquestionably established that globular clusters (GCs) host multiple stellar populations (MPs).
Compelling evidence collected during the years include multiple evolutionary sequences in their colour-magnitude diagrams (see e.g., \citealt{milone2017} and references therein), along with chemical inhomogeneities as depicted from spectroscopic investigations (e.g., \citealt{carretta2009}, \citeyear{carretta2018}; \citealt{marino2015} ).
The lesson we learnt from large sample of stars  analysed in terms of their proton-capture element content (C, N,O, Na, Mg, Al) is that virtually all the Galactic GC stars exhibit the so-called Na-O anti-correlation, although with different shape and extent depending on cluster's mass and metallicity (and a combination of both, we refer to
\citealt{gratton2019} for an updated review). This peculiar chemical pattern seems to suggest that different episodes of star formation occurred in the early stages of GCs: a fraction of the first-generation (FG) stars have activated in their interiors hot H burning (via CNO cycle) providing the processed material from which subsequent generations of stars have been forged (topical papers in this field include \citealt{renzini2008}, \citealt{dantona2016}, \citealt{bastian2018}). The fundamental issue is that we are not able to identify the nature of intra-cluster polluters in GCs: 
principal candidates contain intermediate-mass AGB (IM-AGB) stars during hot-bottom burning (\citealt{ventura2001}), fast-rotating massive single or binary stars (\citealt{decressin2007}, \citealt{demink2009}) and super massive stars (\citealt{denissenkov2014}). Lithium grants a key diagnostics in this context. In fact, it is expected that at CNO (and NeNa) cycle temperatures, no Li is left (Li burns at 
T$\approx$2.5 MK), so that polluting materials should have Li $\sim$ 0 (under the assumption that there is no Li production within the polluters). In this simple scenario Na-poor, O-rich stars (FG stars) should be Li-rich (at the Spite plateau level, in compliance with field halo stars), whereas Na-rich, O-poor stars (second-generation, SG stars) should be Li poor. Thus, Li and O should be positively correlated, while Li and Na anti (or negative) correlated. 
As we will see in the following Section 2, observations do not validate this quite simple theoretical expectation.
Most interesting, while binary or fast-rotating massive stars can only destroy Li, the IM-AGB stars can also produce it via the $^7$Be transport mechanism \citep{cameron1971}.
\section{Observational evidence}
\begin{figure}[h!]
{\includegraphics[width=0.49\textwidth]{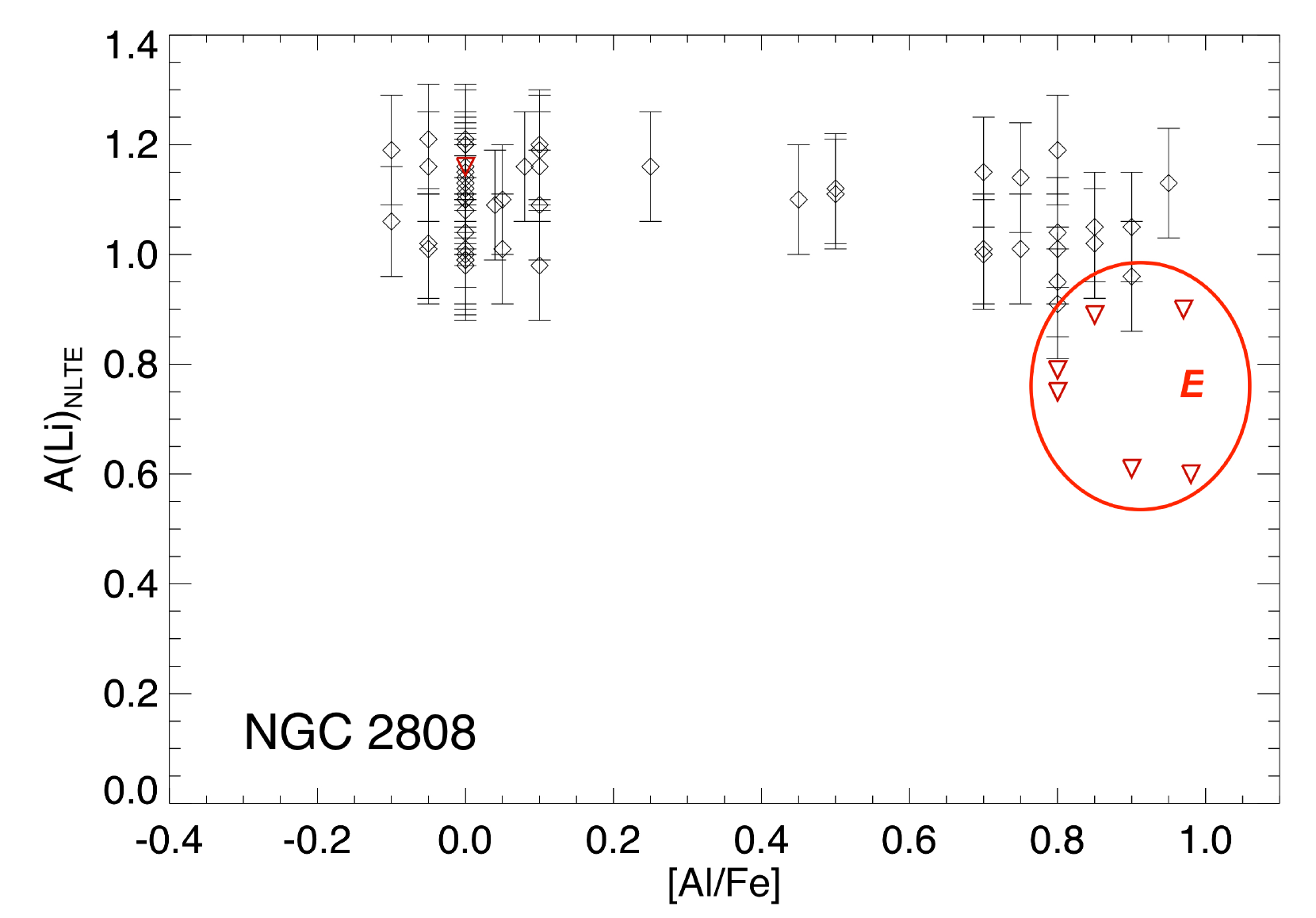}}
\caption{Li abundances as a function of [Al/Fe] for GC giants (below the bump luminosity) in NGC 2808. The figure is taken from \cite{gratton2019}.}
\label{fig:ngc2808}
\end{figure}

Currently, simultaneous measurements for Li and at least one element involved in p-capture reactions are available for 11 GCs (both dwarfs and giants below the bump), including the very well-studied clusters like NGC 6121 (M4, \citealt{dm10}, \citealt{mucciarelli11}), NGC 6397 \citep{lind2009}, and peculiar clusters such as e.g., 47 Tuc \citep{dobrovolskas2014} and $\omega$ Centauri \citep{mucciarelli2018}.
We refer the reader to our review \citep{gratton2019} for the complete list of GCs and corresponding references.
Results indicate that in clusters like M4 (but also M12, NGC 362; \citealt{dorazi2015}) SG stars share the same Li abundance of FG stars: thus, the run of Li with Na/O abundances 
is flat, with no positive or negative correlations between these species. A very weak hint of Li-Na anti-correlation (although the trend is actually driven by only three stars) 
has been obtained in NGC 6397 by \cite{lind2009}; nonetheless, the majority of SG stars are significantly enriched in Li. On the other hand, a positive correlation between Li and O seems to emerge from the analysis of GC dwarfs in NGC 6752 (\citealt{shen2010}), but the slope is not 1: this confirms the presence of stars enriched in Li and depleted in O. 
{\em In general, all spectroscopic investigations carried out so far converge towards the need for Li production in all the GCs. At the time of writing, the only stellar source capable to 
reproduce such pattern are IM-AGB stars}\footnote{It is noteworthy in this context the new study presented at this meeting by D. Sz\'ecsi, which is focussed on the Li production in blue super-giant stars.} Very interesting, the Li behaviour in conjunction with p-capture elements is observed to vary from cluster to cluster. The production seems to be very efficient in relatively small clusters (as M4, where the expected anti-correlations are totally erased from the Li production within the polluters); conversely in massive clusters like NGC 2808 a different picture is emerging. In Fig.~\ref{fig:ngc2808}, [Al/Fe] ratios are shown as a function of Li abundances
for GC giants:  the dominant population of SG stars is still very rich Li in lithium. However, there is an exiguous number of extreme stars 
(labelled as E in the figure, 18$\pm$4 \% of the total GC population) that display Li depletion at some level. For this population we cannot conclude whether the polluters 
were massive stars or a different sub-class (perhaps different range of mass and metallicity) of AGB stars, for which Li production were not effective.
The outcome is that in this GC (and in a much more extreme fashion also in $\omega$ Centauri) not a single polluter source is at work: independently the same conclusion is reached from other chemical tracers (see references in \citealt{gratton2019}).

\section{Discussion and concluding remarks} 
\begin{figure}[h!]
{\includegraphics[width=0.52\textwidth]{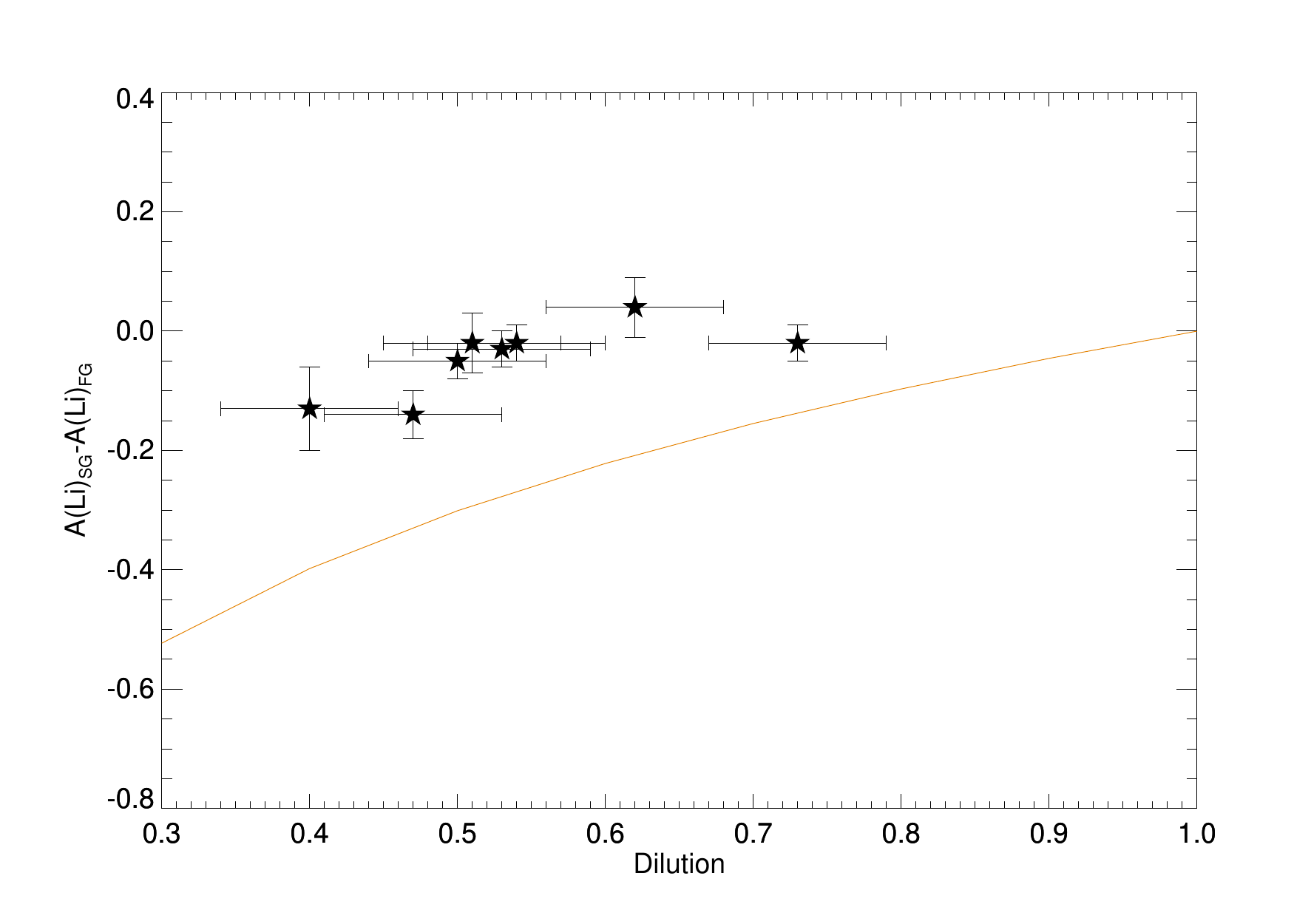}}
\caption{Differences in Li abundances between first and intermediate second-generation stars as a function of dilution factors for a sample of clusters (from \citealt{gratton2019}).}
\label{fig:dilution}
\end{figure}

By considering the observational results collected so far as for Li abundances in SG stars, it is suddenly evident a sort of conspiracy: the Li production has to be exactly at the same level of FG stars, in order to reproduce the flatness of the Li distributions; the proper amount might depends on the source of diluting material (if this is pristine or contaminated by e.g., other stellar source like interacting binaries complicates further the picture).  
In Fig.~\ref{fig:dilution}
we plot the difference in Li abundances for FG and SG stars (only intermediate population, i.e., no extreme population) as a function of the dilution factors (where dilution 0 means pure ejecta and dilution 1 is for pristine material; see \cite{carretta2018} and references therein). The curve is the dilution under the assumption that there is no Li production within the polluters: as it can be seen from the plot, this curve is systematically lower than the observed point indicating that we need to produce lithium within the polluters.
Despite being a quite strong constraint to disentangle the polluter class of GC, lithium abundances have been mostly overlooked so far because either some polluter sources are not able to produce it, or even within the AGB scenario many uncertainties plague the modelling (treatment of convection, mass loss rate, nuclear reaction rates). 
We believe that no model can be considered as reliable in attempting to explain the multiple population scenarios in GC, without taking into account Li abundances.

\begin{acknowledgements}
VD acknowledges support from COST Action CA16117 during this meeting. 
We thank E. Carretta, A. Bragaglia, and S. Lucatello for useful discussions on this topic.
\end{acknowledgements}

\bibliographystyle{aa}

\end{document}